\documentclass{article}

\newtheorem{theorem}{Theorem}

\begin{document}

\title{Strong cosmic censorship for solutions of the Einstein-Maxwell 
field equations with polarized Gowdy symmetry.}

\author{Ernesto Nungesser and Alan D. Rendall \\ 
Max-Planck-Institut f\"ur Gravitationsphysik\\
Albert-Einstein-Institut\\
Am M\"uhlenberg 1\\
14476 Potsdam, Germany}

\date{}

\maketitle

\begin{abstract}
A proof of strong cosmic censorship is presented for a class of solutions 
of the Einstein-Maxwell equations, those with polarized Gowdy symmetry.
A key element of the argument is the observation that by means of a suitable
choice of variables the central equations in this problem can be written
in a form where they are identical to the central equations for general
(i.e. non-polarized) vacuum Gowdy spacetimes. Using this it is seen that 
the deep results of Ringstr\"om on strong cosmic censorship in the 
vacuum case have implications for the Einstein-Maxwell case. Working out
the geometrical meaning of these analytical results leads to the main
conclusion.
\end{abstract}

\section{Introduction}\label{intro}
The Gowdy spacetimes are a class of solutions of the vacuum Einstein equations
with two commuting spacelike Killing vectors. Factoring by the symmetry leads 
to effective equations in one time and one space dimension. To date the Gowdy
spacetimes are one of the best model systems for obtaining a mathematical 
understanding of the dynamics of inhomogeneous solutions of the Einstein
equations. It is known that close analogues of the Gowdy spacetimes exist
in the case that the vacuum Einstein equations are replaced by the 
source-free Einstein-Maxwell equations (see for instance \cite{carmeli}). It 
is thus natural to ask to what extent the known results on the physical 
properties of 
Gowdy spacetimes can be generalized to the case with an electromagnetic field. 
This question is the subject of this paper. When referring to Gowdy spacetimes 
in the following the spatial topology is always assumed to be $T^3$ (a 
three-dimensional torus) except where the contrary is stated explicitly.

A variety of deep mathematical results have been obtained concerning 
Gowdy spacetimes, culminating in the work of Hans Ringstr\"om 
\cite{ringstrom04}, \cite{ringstrom06a}, \cite{ringstromta}. In particular
it was shown in \cite{ringstromta} that strong cosmic censorship holds 
in the class of Gowdy spacetimes. This means that for generic initial 
data within this class the corresponding maximal globally hyperbolic
development is inextendible. (Precise definitions of these terms are
given later in the paper.) A relatively simple special case of the Gowdy
spacetimes are those which are polarized. In that case the main field 
equation becomes linear and this simplifies many things. In the case
of the Einstein-Maxwell equations it is also possible to make a definition of
polarized Gowdy symmetry but in that case the main field equations are
no longer linear. In fact, as shown explicitly below, when expressed in 
suitable variables they are identical to the full (non-polarized) Gowdy
equations of the vacuum case. Hence the results of Ringstr\"om concerning 
the solutions of the partial differential equations can be taken over 
directly. On the other hand, the geometrical and physical meaning of the 
variables is different in the two cases so that the interpretation of the 
results is different. This will be worked out in detail in the following.
It leads in particular to a result on strong cosmic censorship for 
solutions of the Einstein-Maxwell equations with polarized Gowdy symmetry
(Theorem \ref{scc}).

To put this result into context it is worth recalling for which matter
models strong cosmic censorship has been proved for any inhomogeneous
solutions of the Einstein-matter equations. In \cite{dafermos06a} strong
cosmic censorship was proved for solutions of the Einstein-Vlasov system
with $T^2$ symmetry. The genericity assumption made there excludes the
vacuum case. The matter content is used in an essential way. Although
it is plausible that the methods of \cite{dafermos06a} could be extended 
to many other matter models they do not suffice to treat the case of a 
Maxwell field alone. Similar results have been proved for solutions of the 
Einstein-Vlasov system with spherical symmetry and hyperbolic symmetry in 
\cite{dafermos07a}. The results of \cite{dafermos06a} were extended to the 
case with positive cosmological constant in \cite{smulevici}. Strong cosmic 
censorship has been proved for the plane symmetric Einstein-scalar field 
system in \cite{tegankong}. For those Gowdy spacetimes (not considered 
further in the following) with topologies other than $T^3$ cosmic 
censorship has been studied in the polarized case in \cite{chrusciel90}.
The results which have been listed are those existing positive results 
known to the authors on strong cosmic censorship of this level of generality.
They are valid for inhomogeneous solutions of the field equations, with or 
without matter. They concern solutions which belong to some symmetry class 
but do not require smallness assumptions on the initial data. 

Apart from cosmic censorship, more detailed information is obtained about 
the qualitative properties of the solutions of the Einstein-Maxwell equations
near the initial singularity and at late times. For generic solutions it is 
shown that the Kretschmann scalar 
$R_{\alpha\beta\gamma\delta}R^{\alpha\beta\gamma\delta}$ blows up uniformly on
certain regions as the singularity is approached. It is also shown that the 
spacetimes are geodesically complete towards the future and that they 
exhibit oscillatory behaviour at late times analogous to that found in
\cite{ringstrom04} for Gowdy spacetimes.

In Section \ref{symm} definitions of Gowdy and polarized Gowdy symmetry are 
introduced. These definitions, which make use of discrete symmetries, have
not previously been worked out in the literature and so are treated in 
detail here. In Section \ref{basic}, a number of equations are displayed. 
They are the basic equations of the class of spacetimes which are of interest 
in the sequel. In Sections 
\ref{singularity} and \ref{future} the main results on the asymptotic 
behaviour of solutions in the past and future direction respectively are 
proved. These results are combined in Section \ref{conclusion} to give the 
main result (Theorem \ref{scc}) and some thoughts are presented on how the 
results of this paper might be generalized in the future.

This paper is based in part on the diploma thesis of the first author 
\cite{nungesser}. 

\section{Gowdy spacetimes and discrete symmetries}\label{symm}

Let $g_{\alpha\beta}$ be a Lorentzian metric on a four-dimensional manifold 
$M$ and suppose that $g_{\alpha\beta}$ is a solution of the Einstein equations 
coupled to some matter model. A symmetry of this solution is an isometry of 
the metric which also preserves the matter fields. The solution is said to 
have $T^2$ (or $U(1)\times U(1)$) symmetry if this group acts on $M$ by 
symmetries of the solution. Think of a solution as being defined by Cauchy 
data. If the Cauchy data are invariant under a symmetry group and the Einstein
equations coupled to the given matter model have a well-posed initial value
problem then there is an associated action of the group on the solution
(cf. the discussion in section 9.2 of \cite{rendall08}). 

Now suppose a spacetime is given where the metric takes the form
\begin{equation}\label{t2}
e^{2(\eta-U)}(-dt^2+d\theta^2)+e^{2U}[dx+Ady+(G+AH)d\theta]^2+
e^{-2U}R^2[dy+Hd\theta]^2.
\end{equation}
Here $\theta$, $x$ and $y$ are periodic coordinates on $T^3$ and the functions
$\eta$, $U$, $A$, $G$, $H$ and $R$ depend only on $t$ and $\theta$. This 
metric has $T^2$ symmetry with the group action generated by the Killing 
vectors $\frac{\partial}{\partial x}$ and $\frac{\partial}{\partial y}$. This 
is a form of the metric used in \cite{arw}. In that paper it was shown that
in any globally hyperbolic spacetime with $T^2$ symmetry and a Cauchy 
hypersurface diffeomorphic to $T^3$ coordinates of this type can be 
introduced in a neighbourhood of any Cauchy hypersurface. 

A metric of the form (\ref{t2}) will be said to have Gowdy symmetry if the 
transformation which simultaneously maps $x$ to $-x$ and $y$ to $-y$ is an
isometry. It follows that $G$ and $H$ must vanish.  Conversely, if $G$ and 
$H$ vanish the spacetime has this extra symmetry. The terminology comes
from the fact that the solutions of the vacuum Einstein equations with this
symmetry are precisely the Gowdy spacetimes. A metric with Gowdy 
symmetry is said to be polarized if the individual transformations mapping 
$x$ to $-x$ and $y$ to $-y$ are symmetries. This is equivalent to the 
vanishing of $A$. Next a more abstract characterization in terms of discrete
symmetries will be derived. Let $G_1$ be the group of isometries of ${\bf R}^2$
generated by translations and reflection in the origin. Let the latter 
transformation be denoted by $R_{1,2}$. $G_1$ is a semidirect product of 
${\bf R}^2$ and $Z_2$. Suppose that in a spacetime with $T^2$ symmetry the 
action of $T^2$ extends to an action of $G_1$ by isometries. Then the 
coordinate form of the transformation corresponding to $R_{1,2}$ is given up to 
an overall translation by the transformation which simultaneously sends $x$ to 
$-x$ and $y$ to $-y$. This follows from the relation 
$R_{1,2}T_{a,b}=T_{-a,-b}R_{1,2}$ which holds for 
any translation $T_{a,b}$. To see this, note first that the isometry 
corresponding to $R_{1,2}$ cannot be a translation. Hence it has a fixed point. 
Evaluating the relation at that fixed point gives the desired conclusion for
the restriction to one orbit. Consider the geodesic in $S$ through that point
orthogonal to the orbit. Its intersection with any other orbit must also be
a fixed point of $R_{1,2}$. Thus the necessary translation is the same for
all orbits on the initial hypersurface. The same conclusion can then be 
obtained for any orbit in the spacetime by a similar argument using a 
geodesic of the spacetime through the fixed point which is orthogonal to
the initial hypersurface. In this way an abstract characterization of Gowdy 
symmetry in terms of an action of the group $G_1$ is obtained. Similarly, 
polarized Gowdy symmetry can be characterized by the fact that the $T^2$ 
action extends to an action by isometries of the group $G_2$ of isometries of 
${\bf R}^2$ generated by translations and the reflections $R_1$ and $R_2$ in 
the two Cartesian coordinates. The group $G_2$ is a semidirect product of 
${\bf R}^2$ with $Z_2\oplus Z_2$. The elements $R_1$ and $R_2$ satisfy the 
relations $R_1T_{a,b}=T_{-a,b}R_1$ and $R_2T_{a,b}=T_{a,-b}R_2$. It follows by 
the same type of argument as in the previous case that $R_1$ and $R_2$ can be 
identified with reflections in $x$ and $y$. 

Let us suppose for simplicity that the torus has unit coordinate volume with 
respect to the coordinates $(x,y)$. Then the area of an orbit of the symmetry 
group contained in the domain of the above coordinate system is given by 
$R$. This can be used to extend the definition of $R$ to the whole spacetime 
by requiring its value at a given point to be equal to the  
area of the orbit on which that point lies. The function $R$ is called the 
area radius. It will be assumed in the following that the Maxwell field
does not vanish identically so that the spacetime is not flat. Recall that
the Maxwell field satisfies the dominant energy condition. A proof of this is
given in the Appendix. It follows from Proposition 3.1 of \cite{rendall97a} 
that the gradient of $R$ is always timelike. As a consequence the function $R$ 
can be used as a time coordinate. This is referred to as an areal time 
coordinate. It is not a priori clear whether in a given solution there is any 
level surface of $R$ which is a Cauchy surface. An argument which shows that
there is a level surface of this kind is presented in the next section.

Having defined Gowdy and polarized Gowdy symmetry for metrics the next
step is to do the same for solutions of the Einstein-Maxwell equations.
To do this the action of the groups $G_1$ and $G_2$ must be extended to
the Maxwell field. Any diffeomorphism transports any tensor field in
a natural way. If the diffeomorphism is an isometry consistency with the 
Einstein equations requires that the energy-momentum tensor must be 
transported in this way. It is, however, possible to transport matter fields
in a different way and this will be done here. Consider the case that the 
electromagnetic field is defined by a four-potential $A_\alpha$. If $\phi$ 
is a diffeomorphism consider the transformation 
$(g,A)\mapsto (\phi_*g,-\phi_*A)$. This leads to the transformation 
$F\mapsto -\phi_* F$ for the electromagnetic field. This ensures that the 
energy-momentum tensor is left invariant. If the action of the group
$G_1$ is extended so as to transform the electromagnetic field in this 
way then $A_0$ and $A_1$ vanish. Denote the remaining components of
the potential by $\omega=A_2$ and $\chi=A_3$. This provides the 
definition of Gowdy symmetry for the Einstein-Maxwell system. The only
non-trivial components of the electromagnetic field are $F_{02}$,
$F_{03}$, $F_{12}$ and $F_{13}$. Taking instead $G_2$ provides the definition 
of polarized Gowdy symmetry. This does not lead to a direct further 
restriction on the potential but does further restrict the metric. It will be 
shown in the next section that the Einstein-Maxwell equations lead to a 
consistency condition which apparently forces one of the components of the 
potential to be set to zero in the polarized case.  

It would have been possible to use the more straightforward transformation
$(g,A)\mapsto (\phi_*g,\phi_*A)$ instead of that just introduced. This leads
to a different class of solutions of the Einstein-Maxwell equations. In
that case the only non-trivial components of the potential are $A_0$ and
$A_1$. The only component of the electromagnetic field which may be 
non-vanishing in this case is then $F_{01}$. If fields are allowed which
do not arise from a potential then $F_{23}$ may also be non-vanishing. 
Solutions of the Einstein-Maxwell equations of this type were considered in 
\cite{weaverphd}. It seems that they exhibit oscillatory behaviour near the
singularity and are thus more complicated than the solutions considered in
the following. This other symmetry class is not considered further in 
the present work.  

\section{Basic equations}\label{basic}

Consider Cauchy data for a solution of the Einstein-Maxwell equations with
polarized Gowdy symmetry and suppose that these data have a constant value
of $R$. In the solution evolving from these data coordinates exist on a 
neighbourhood of the initial hypersurface where the metric takes the form
(\ref{t2}) with $G=H=A=0$. The function $R$ can be introduced as a new time 
coordinate and after that a new spatial coordinate can be introduced so that 
$g_{01}=0$. Then the transformed metric is of the form (\ref{t2}) with 
$G=H=A=0$ and $R=t$ except for the fact that $dt^2$ may have been replaced 
by an expression of the form $\alpha dt^2$ for some function 
$\alpha (t,\theta)$. Using the fact that for the Maxwell field $T^0_0=-T^1_1$ 
one of the field equations (equation (11.42) on p. 200 of \cite{rendall08}) 
implies that $\alpha_t=0$. It is possible, while maintaining the coordinate 
conditions imposed up to now, to reparametrize the coordinate $\theta$ on the 
initial hypersurface. Because $\alpha$ does not depend on time this can be 
used to impose the condition $\alpha=1$. Introduce new variables as follows:
$\lambda=4(\eta-U)+\log t$, $P=2U-\log t$.  Then the metric takes the form
\begin{equation}\label{gowdymetric}
t^{-\frac12}e^{\frac{\lambda}2}(-dt^2+d\theta^2)+t(e^Pdx^2+e^{-P}dy^2)
\end{equation}
In the new coordinates the original Cauchy hypersurface is not a hypersurface
of constant $t$. That there does exist a Cauchy hypersurface of constant $t$
can be shown as follows. The maximal Cauchy development of the initial data
corresponds to a certain subset of $(t,\theta)$-space. Suppose that this set
has a past boundary point $p$ at which $t$ is strictly positive. By using 
standard light cone estimates for the Gowdy equations it can be shown that 
$P$, $\chi$ and all their derivatives all orders are uniformly bounded on the 
intersection of the region corresponding to the maximal Cauchy development 
with a small neighbourhood of $p$. By uniform continuity they extend 
continuously to $p$. Using arguments as in sections 4 and 5 of \cite{arr} or
as in \cite{bcim} it can be concluded that the solution can be extended to
a larger region in such a way as to define a globally hyperbolic extension.
This contradicts the definition of the maximal Cauchy development. Hence 
no such boundary point $p$ exists.  The solution exists all the way to $t=0$
and hence if $t_0$ is any positive number smaller than the minimum of $t$
on the original Cauchy hypersurface then $t=t_0$ will be a Cauchy surface
of the type whose existence was to be demonstrated. 

As explained in the previous section the only non-vanishing components
of the four-potential defining the Maxwell field are $A_2=\omega$ and 
$A_3=\chi$. With these choices the Einstein-Maxwell equations can be computed.
The details are given in \cite{nungesser}.
\begin{eqnarray}
&&-P_{tt}-t^{-1}P_t+P_{\theta\theta}=-2t^{-1}[e^{-P}(-\omega_t^2+\omega_\theta^2)
-e^P(-\chi_t^2+\chi_\theta^2)]\label{gowdy1}     \\
&&-\omega_{tt}+\omega_{\theta\theta}=
-P_t\omega_t+P_\theta\omega_\theta\label{gowdy2}   \\
&&-\chi_{tt}+\chi_{\theta\theta}=
-[-P_t\chi_t+P_\theta\chi_\theta]\label{gowdy3}   \\
&&\lambda_t=t(P_t^2+P_\theta^2)+4[e^{-P}(\omega_t^2+\omega_\theta^2)
+e^P(\chi_t^2+\chi_\theta^2)]       \label{gowdy4}   \\
&&\lambda_\theta=2tP_tP_\theta+8(e^{-P}\omega_t\omega_\theta
+e^P\chi_t\chi_\theta)               \label{gowdy5}   \\
&&\omega_t\chi_t=\omega_\theta\chi_\theta\label{gowdy6}
\end{eqnarray}
The full set of Einstein equations includes one involving the second
derivatives of $\lambda$. It is a consequence of (\ref{gowdy4}) and 
(\ref{gowdy5}) and plays a minor role in what follows. It is written
down in terms of other variables later. The equation 
(\ref{gowdy6}) is the consistency condition alluded to in the last 
section. The only large classes of solutions known which are consistent with
this equation are those where either $\omega$ or $\chi$ vanishes. In 
\cite{carmeli} the authors considered other possibilities but this did not 
lead to the discovery of other interesting classes of solutions. 
In most of what follows it will be assumed that $\omega=0$. Assuming
$\chi=0$ would lead to an equivalent problem. These two cases are related 
by interchanging the roles of the two Killing vectors.   

The equations (\ref{gowdy1}), (\ref{gowdy3}), (\ref{gowdy4}) and 
(\ref{gowdy5}) with $\omega=0$ can be shown to be equivalent to the equations 
for (not necessarily polarized) Gowdy vacuum spacetimes by introducing new
variables. These are given by
\begin{eqnarray}
&&\bar P=\frac12 (P-\log t)      \label{gowdyvac1} \\
&&\bar\lambda=\frac14(\lambda-\log t)-\bar P           \label{gowdyvac2} 
\end{eqnarray}  
Then the equations for $\bar P$, $\chi$ and $\lambda$ coming from
(\ref{gowdy1}), (\ref{gowdy3}), (\ref{gowdy4}) and (\ref{gowdy5})
are the basic equations for Gowdy vacuum spacetimes (cf. (\cite{ringstrom04})
with $P$ replaced by $\bar P$, $Q$ replaced by $\chi$ and $\lambda$
replaced by $\bar\lambda$. These equations are:
\begin{eqnarray}
&&-\bar P_{tt}-t^{-1}\bar P_t+\bar P_{\theta\theta}=
e^{2\bar P}(-\chi_t^2+\chi_\theta^2)\label{gowdy7}                          \\
&&-\chi_{tt}-t^{-1}\chi_t+\chi_{\theta\theta}
=-2(-\bar P_t\chi_t+\bar P_\theta\chi_\theta)\label{gowdy8}                 \\
&&\bar\lambda_t=t[\bar P_t^2+\bar P_\theta^2+e^{2\bar P}
(\chi_t^2+\chi_\theta^2)]\label{gowdy9}                                     \\
&&\bar\lambda_\theta=2t[\bar P_t\bar P_\theta+e^{2\bar P}\chi_t\chi_\theta]
\label{gowdy10}    
\end{eqnarray} 
In terms of these variables the metric is
\begin{equation}\label{gowdymetric2}
e^{2(\bar\lambda+\bar P)}(-dt^2+d\theta^2)+t^2e^{2\bar P} dx^2+e^{-2\bar P}dy^2.
\end{equation}
The additional equation involving the second derivative of $\bar\lambda$
reads
\begin{equation}\label{gowdy11}
-\bar\lambda_{tt}-t^{-1}\bar\lambda_t+\bar\lambda_{\theta\theta}
=2(\bar P_\theta^2+e^{2\bar P}\chi_\theta^2).
\end{equation}

\section{Structure of the singularity}\label{singularity}

Since the equations for solutions of the Einstein-Maxwell system with
polarized Gowdy symmetry are identical to the well-known Gowdy equations
in the vacuum case it can be concluded from a theorem of \cite{moncrief81}
that the solution corresponding to
any initial data with constant $R$ exists globally on the interval $(0,\infty)$
when expressed in areal time. The purpose of this section is to investigate
the asymptotics of these solutions in the limit $t\to 0$. 
Interpreting a result of \cite{ringstrom06a} in 
terms of the present problem shows that for a fixed value of $\theta$
the quantity 
$t[(\bar P_t^2+e^{2\bar P}\chi_t^2)(t,\theta)]^{\frac12}$
converges to a limit $v_\infty(\theta)$ as $t\to 0$. This quantity is called 
the asymptotic velocity at $\theta$. It turns out to be useful to examine 
regions ${\cal D}_{\theta_0}$ defined as the set of points $(t,\theta)$ such 
that $|\theta-\theta_0|\le t$. Intuitively this can be thought of as the 
causal future of a point of the singularity defined by $\theta=\theta_0$. 

The structure of the singularity will be described in detail for a generic
set ${\cal G}_c$ of initial data, which is strictly analogous to a similar set 
introduced in \cite{ringstrom06a}, using the relation between vacuum solutions 
and polarized Einstein-Maxwell solutions. By analogy with \cite{ringstrom06a}
we can define what is meant by a solution having a non-degenerate false spike 
or a non-degenerate true spike
at some value of $\theta$. The detailed definition will not be reproduced 
here. For convenience call a point $\theta$ an ordinary point for a given 
solution if $0<v_\infty(\theta)<1$ and 
$\lim_{t\to 0}tP_t(t,\theta)=-v_\infty (\theta)$. Define ${\cal G}_{l,m,c}$ to be 
the set of initial data determining solutions for which all points $\theta$ are 
ordinary except for $l$ non-degenerate false spikes and $m$ non-degenerate 
true spikes. Then ${\cal G}_c$ is 
the union of ${\cal G}_{l,m,c}$ over all non-negative integers $l$ and $m$.
Data belonging to ${\cal G}_c$ are called generic. This subset is
open in the set of initial data in the $C^0\times C^1$ topology and dense in 
the $C^\infty$ topology 
\cite{ringstromta}. A comment should be made on the subscript $c$ which is 
included here for 
consistency with Ringstr\"om's notation. It is a reminder that initial data
for the Gowdy equations must satisfy an integral constraint for topological
reasons. An analogous constraint occurs in the Einstein-Maxwell case. It 
arises by integrating equation (\ref{gowdy10}) with respect to $\theta$.
The integral of the left hand side vanishes due to periodicity and so the 
integral of the right hand side must vanish too. If there is a false spike
at $\theta$ then $0<v_\infty(\theta)<1$ while if there is a true spike
$1<v_\infty(\theta)<2$. Thus, in particular, generic data are such that 
$v_\infty$ is never equal to one at any point. In the vacuum case this 
statement is enough to imply blow-up of the Kretschmann scalar along any 
past-directed causal curve due to Proposition 1.19 of \cite{ringstrom06a}.

To prove inextendibility it will be necessary to consider the case of 
normal points, false spikes and true spikes separately. First the case
of a normal point will be examined. The expression for the Kretschmann scalar 
is unwieldy and it might be thought that it would be easier to prove 
inextendibility in the electromagnetic case by showing that the invariant 
$F_{\alpha\beta}F^{\alpha\beta}$ blows up. Unfortunately this turns out not
to be the case. There is a large class of Gowdy solutions (the low velocity 
solutions) which, when translated to the case of interest here, admit 
asymptotic expansions of the following form 
\begin{eqnarray}
&&\bar P(t,\theta)=-v_\infty(\theta)\log t+\ldots\label{lowvelocity1}
  \\
&&\chi(t,\theta)=q(\theta)+\psi (\theta)t^{2v_\infty}+\ldots\label{lowvelocity2}
\end{eqnarray}      
where the terms omitted are lower order as $t\to 0$. It follows that 
$e^{-2\bar\lambda}$ is proportional to $t^{-2v_\infty^2}$ in leading order 
while
\begin{equation}
\chi_\theta^2-\chi_t^2=q_\theta^2-4v_\infty^2 t^{4v_\infty-2}\psi^2+\ldots
\end{equation}
Since
\begin{equation}
F^{\alpha\beta}F_{\alpha\beta}=2e^{-2\bar\lambda}(\chi_\theta^2-\chi_t^2)
\end{equation}
it follows that the information contained in the above expansions and the fact
that $0<v_\infty<1$ do not always suffice to determine whether the invariant 
blows up or not. For if $v_\infty=\frac12$ and $q_\theta^2=4v_\infty^2\psi^2$ the leading 
term in the expansion for $F^{\alpha\beta}F_{\alpha\beta}$ vanishes and whether 
the next to leading term blows up cannot be decided using the available 
information. For the solutions considered here 
$F^{\alpha\beta}{}^*F_{\alpha\beta}=0$ and so this invariant is not helpful.

Next the Kretschmann scalar will be examined. An observation which simplifies
the calculations is that, due to the discrete symmetries of the spacetimes 
under consideration, a component $R_{\alpha\beta\gamma\delta}$ which does not 
vanish must have an even number of indices equal to two and an even number
equal to three. It follows that up to symmetries (not including the 
algebraic Bianchi identity) the only components which may be non-zero are
\begin{equation}
R_{0101}, R_{0202}, R_{0212}, R_{1212}, R_{0303}, R_{0313}, R_{1313}, R_{2323}
\end{equation} 
The Kretschmann scalar is equal to
\begin{eqnarray}
&&4[R_{0101}R^{0101}+R_{0202}R^{0202}+R_{1212}R^{1212}+R_{0303}R^{0303}
\nonumber\\
&&+R_{1313}R^{1313}+R_{2323}R^{2323}]+8[R_{0212}R^{0212}+R_{0313}R^{0313}].
\end{eqnarray}
In terms of the metric coefficients the individual terms are given by
\begin{eqnarray}
R_{0101}R^{0101}&=&t^{-4}e^{-4(\bar\lambda+\bar P)}                   
[-t^2\bar\lambda_{tt}-t^2\bar P_{tt}+t^2(\bar P_{\theta\theta}
+\bar\lambda_{\theta\theta})]^2\nonumber                                     \\
R_{0202}R^{0202}&=&t^{-4}e^{-4(\bar\lambda+\bar P)}                   
[t\bar\lambda_t+t^2\bar P_t\bar\lambda_t-t^2\bar P_{tt}-t\bar P_t
+t^2(\bar P_\theta^2+\bar P_\theta\bar\lambda_\theta)]^2\nonumber            \\
R_{1212}R^{1212}&=&t^{-4}e^{-4(\bar\lambda+\bar P)}                   
[t\bar\lambda_t+t^2\bar\lambda_t\bar P_t+t\bar P_t+t^2\bar P_t^2
-t^2(\bar P_{\theta\theta}-\bar P_\theta\bar\lambda_\theta)]^2\nonumber       \\
R_{0303}R^{0303}&=&t^{-4}e^{-4(\bar\lambda+\bar P)}                   
[t^2\bar P_{tt}-t^2\bar P_t\bar\lambda_t-2t^2\bar P_t^2
-t^2(\bar P_\theta\bar\lambda_\theta+\bar P_\theta^2)]^2\nonumber            \\
R_{1313}R^{1313}&=&t^{-4}e^{-4(\bar\lambda+\bar P)}                   
[-t^2\bar P_t\bar\lambda_t-t^2\bar P_t^2
+t^2(\bar P_{\theta\theta}-2\bar P_\theta^2-\bar P_\theta\bar\lambda_\theta)]^2  
\nonumber\\
R_{2323}R^{2323}&=&t^{-4}e^{-4(\bar\lambda+\bar P)}                   
[-t\bar P_t-t^2\bar P_t^2+t^2\bar P_\theta^2]^2\nonumber                    \\
R_{0212}R^{0212}&=&-t^{-4}e^{-4(\bar\lambda+\bar P)}                   
[-t\bar\lambda_\theta-t^2\bar P_t\bar P_\theta+t^2\bar P_{t\theta}
-t^2\bar P_\theta\bar\lambda_t-t^2\bar\lambda_\theta \bar P_t]^2\nonumber    \\ 
R_{0313}R^{0313}&=&-t^{-4}e^{-4(\bar\lambda+\bar P)}
[-t^2\bar P_{t\theta}+t^2\bar P_t\bar\lambda_\theta
+t^2\bar P_\theta\bar\lambda_t+3t^2\bar P_t\bar P_\theta]^2\nonumber
\end{eqnarray}
The first six terms are manifestly non-negative while the last two are 
manifestly non-positive. The fact that all terms contain a common factor helps 
when comparing their magnitudes. It suffices to compare the magnitudes of
the quantities in square brackets. Suppose it can be shown that in a 
given solution at least one of the positive terms tends to infinity as 
$t\to 0$ and that both the negative terms are neglible with respect to that 
positive term. Then this implies that the Kretschmann scalar tends to infinity.
It turns out, as will be explained in more detail later, that for this 
question the spatial derivatives in these expressions are unimportant 
compared to the time derivatives. The essential information needed for 
the proof is contained in the following expressions: 
\begin{eqnarray}
R_{0101}R^{0101}&=&t^{-4}e^{-4(\bar\lambda+\bar P)}                   
[-t^2\bar\lambda_{tt}-t^2\bar P_{tt}]^2\nonumber+\ldots                            \\
R_{0202}R^{0202}&=&t^{-4}e^{-4(\bar\lambda+\bar P)}                   
[t\bar\lambda_t+t^2\bar P_t\bar\lambda_t-t^2\bar P_{tt}-t\bar P_t]^2+\dots
\nonumber                                                                  \\
R_{1212}R^{1212}&=&t^{-4}e^{-4(\bar\lambda+\bar P)}                   
[t\bar\lambda_t+t^2\bar\lambda_t\bar P_t+t\bar P_t+t^2\bar P_t^2]^2+\ldots
\nonumber                                                                  \\
R_{0303}R^{0303}&=&t^{-4}e^{-4(\bar\lambda+\bar P)}                   
[t^2\bar P_{tt}-t^2\bar P_t\bar\lambda_t-2t^2\bar P_t^2]^2+\ldots\nonumber         \\
R_{1313}R^{1313}&=&t^{-4}e^{-4(\bar\lambda+\bar P)}                   
[-t^2\bar P_t\bar\lambda_t-t^2\bar P_t^2]^2+\dots  
\nonumber                                                                  \\
R_{2323}R^{2323}&=&t^{-4}e^{-4(\bar\lambda+\bar P)}                   
[-t\bar P_t-t^2\bar P_t^2]^2+\dots\nonumber                                       \\
R_{0212}R^{0212}&=&\ldots\nonumber                                            \\ 
R_{0313}R^{0313}&=&\ldots\nonumber
\end{eqnarray}
The terms not explicitly written in these equations are not relevant for
the proof which follows. It is not claimed that in the approach to the 
singularity each of the terms which have been omitted is small in comparison 
to the terms which have been written out explicitly.

Consider now the case of a normal point $\theta_0$. Then Proposition 1.5
of \cite{ringstrom06a} provides information about the limiting behaviour
of the solution in a neighbourhood of $\theta_0$. It implies that 
$t\bar P_t$ converges to $-v_\infty$ and that $t\bar P_\theta$ and 
$t^2\bar P_{t\theta}$ tend to zero. It also shows that if equations 
(\ref{gowdy9}) and (\ref{gowdy10}) are multiplied by $t$ the terms 
containing derivatives of $\chi$
on the right hand sides tend to zero as $t\to 0$. If follows that
$t\bar\lambda_t$ converges to $v_\infty^2$ and $t\bar\lambda_\theta$ tends to
zero as $t\to 0$. These facts together show that the quantities in square
brackets in the expressions for $R_{0212}R^{0212}$ and $R_{0313}R^{0313}$
tend to zero at the singularity. The expression in the square bracket in
$R_{2323}R^{2323}$ converges to $v_\infty-v_\infty^2\ne 0$. The expression 
$e^{-4(\bar P+\bar\lambda)}$ is asymptotic to an expression proportional to
$t^{4(v_\infty-v_\infty^2)}$. Since $v_\infty(1-v_\infty)<1$ this is enough to 
conclude that the curvature blows up uniformly in a neighbourhood of a 
normal point $\theta_0$. It is also part of the conclusions of Proposition 1.5 
of \cite{ringstrom06a} that if $\theta_0$ is a false spike a change of 
variables can be carried out which gives a new solution with the asymptotic 
properties which hold for a normal point. Since it is known that this 
transformation does not change the spacetime geometry but only the way it is 
represented the curvature properties are unaffected. Thus curvature blow-up 
also holds in the case of a false spike. It remains to treat the case of a 
true spike at a point $\theta_0$. In that case Proposition 6.7 of 
\cite{ringstrom06a} implies that $t\bar P_\theta$ and $te^{\bar P}\chi_\theta$ 
converge to zero uniformly on ${\cal D}_{\theta_0}$. Corollary 6.8 shows that 
$t^2P_{t\theta}$ also converges to zero there. By Proposition 6.11 of 
\cite{ringstrom06a} and the remark following it $t\bar P_t$ and 
$te^{\bar P}\chi_t$ converge to $-v_\infty$ and zero respectively on that 
region. In this case $v_\infty(1-v_\infty)<0$. Thus the Kretschmann
scalar can be controlled as in the other cases, with the difference that 
the control is only obtained on the region ${\cal D}_{\theta_0}$ rather than
for an interval of $\theta$ containing $\theta_0$.  

In the following theorem the topology on the set of initial data used is the 
$C^\infty$ topology.

\begin{theorem}\label{blowup}
There is an open dense subset ${\cal G}_c$ of the set of 
smooth initial data for the Einstein-Maxwell equations with polarized Gowdy 
symmetry and constant areal time such that the Kretschmann scalar tends to 
infinity along any inextendible past-directed causal geodesic.
\end{theorem}

\noindent
{\bf Proof} The set ${\cal G}_c$ has been defined above. For a solution
evolving from data belonging to this set each point $\theta_0$ is either
an ordinary point, a false spike or a true spike. It has been shown that
in each of these three cases the Kretschmann scalar blows up uniformly
on ${\cal D}_{\theta_0}$ as $t\to 0$. It is shown in \cite{ringstrom06a}
that any inextendible past-directed causal geodesic eventually lies in a
region of this type with some value of $\theta_0$ as $t\to 0$. The 
proof of this is pure Lorentzian geometry and makes no use of the Einstein 
equations. Hence it also applies to the situation of this theorem. 

\section{Future geodesic completeness}\label{future}

This section is concerned with the asymptotic behaviour of solutions in the 
future, i.e. with the behaviour as the areal time $t$ tends to infinity.
The asymptotic behaviour of solutions of the vacuum Gowdy equations has been
determined in detail in \cite{ringstrom04}. This information suffices in 
particular to prove future geodesic completeness in the vacuum case. It will be 
shown that similar arguments can be used to give a proof of future geodesic 
completeness of solutions of the Einstein-Maxwell equations with polarized
Gowdy symmetry.  

As in the vacuum case it is convenient in the proof of geodesic completeness
to distinguish between homogeneous and inhomogeneous spacetimes. The 
homogeneous case is that in which $\bar P$, $\chi$ and $\bar\lambda$ depend 
only on $t$. Geodesic completeness in the homogeneous case follows from Theorem
2.1 of \cite{rendall95}. That theorem applies to solutions of the Einstein
equations coupled to a general class of matter models which satisfy two
conditions, the dominant energy condition and the non-negative sum pressures
condition. Both of these hold for the Maxwell field. In fact the second 
follows from the first because the energy-momentum tensor of the Maxwell field 
has vanishing trace. It should also be noted that for these solutions, which 
are of Bianchi type I, the mean curvature of the hypersurfaces of constant $t$ 
is everywhere negative and hence, in particular, bounded above.

It remains to consider the inhomogeneous case. From Proposition 1.8 of 
\cite{ringstrom04} it can be concluded that $\bar P_t$, $\bar P_\theta$,
$e^{\bar P}\chi_t$ and $e^{\bar P}\chi_\theta$ are $O(t^{-1/2})$ as $t\to\infty$,
uniformly in $\theta$. This means in particular that $\bar\lambda_t$ is 
bounded in the future. From Theorem 1.7 of \cite{ringstrom04} it follows that
$\bar\lambda=c_{\bar\lambda}t+O(\log t)$ for a positive constant 
$c_{\bar\lambda}$ as $t\to\infty$. It is at this point that the condition that 
the solution is inhomogeneous is used. Combining these facts shows that
$`<\bar\lambda+\bar P=c_{\bar\lambda}t+O(t^{\frac12})$. Define an orthonormal 
frame $\{e_\mu\}$ by normalizing the coordinate basis vectors 
$\frac{\partial}{\partial t}$, $\frac{\partial}{\partial \theta}$, 
$\frac{\partial}{\partial x}$, $\frac{\partial}{\partial y}$. A
computation shows that the corresponding rotation coefficients
satisfy an inequality of the form
\begin{equation} 
|g(\nabla_{e_\mu}e_\nu,e_\rho)|\le Ct^{-\frac12}e^{-\bar\lambda-\bar P}
\end{equation}
for a constant $C$ whenever at least two of the indices $(\mu,\nu,\rho)$ 
are equal to two or three. The facts just listed suffice to show that the 
argument used to prove geodesic completeness in the vacuum case in 
\cite{ringstrom04} also applies to the Einstein-Maxwell case, proving 
geodesic completeness in that case too.

In \cite{ringstrom04} further information about the late-time behaviour 
of Gowdy spacetimes is derived. It is shown that, remarkably, there are
solutions with oscillations of $P$ which persist for all times. More
precisely, the amplitude of the oscillation does not decay and the fact 
that $P_t$ converges to zero is explained by the fact that the frequency
of the oscillation decreases with time. The derivative $P_\theta$ does 
decay, which means that in a sense the solution becomes homogeneous
asymptotically. Nevertheless the dynamics of these solutions is different
from that of any homogeneous solution. These solutions can be used to
produce solutions of the Einstein-Maxwell equations by the correspondence
which is central to this paper. The statements made about $P$ in the vacuum
case apply directly to $\bar P$ in the electromagnetic case. Considering
$\chi$ it is seen that the electromagnetic potential oscillates with an
amplitude which does not decay while the decrease in the frequency of the 
oscillation leads to decay of the electric field. The potential becomes 
spatially homogeneous asymptotically and this gives decay of the magnetic 
field.    

A key step in proving the results about future asymptotics in the vacuum case
is to show that the following energy-like quantity tends to zero as 
$t\to \infty$. 
\begin{equation}\label{energy1} 
{\cal E}_1 (t)=\frac12 \int_{S^1} [P_t^2+P_\theta^2+e^{2P}(Q_t^2+Q_\theta^2)] 
(t,\theta) d\theta. 
\end{equation}
Of course there is a corresponding quantity in the Einstein-Maxwell case
which also tends to zero as $t\to\infty$. It is given explicitly by
\begin{equation}\label{energy2} 
{\cal E}_2 (t)=\frac12\int_{S^1} [\bar P_t^2+\bar P_\theta^2+e^{2\bar P}
(\chi_t^2+\chi_\theta^2)] 
(t,\theta) d\theta. 
\end{equation}

\section{Conclusions and outlook}\label{conclusion} 

Given a matter model in general relativity for which the Einstein-matter
equations have a well-posed initial value problem there exists a maximal
Cauchy development corresponding to each initial data set. The strong 
cosmic censorship hypothesis says that for generic initial data this
maximal development is not extendible to a larger spacetime. There are
some technical choices involved in this definition: how much regularity
is assumed for the data, what topology is put on these data and what
regularity properties are assumed of an extension. Here no general
discussion of these issues will be given - we will just make one reasonable
choice. It is assumed that the data are smooth, i.e. $C^\infty$. They are
topologized using the $C^\infty$ topology, i.e. the topology of uniform
convergence of a geometrical object together with its derivatives of all
orders. Extendibility is taken in the sense of extensions of the geometry
of class $C^2$. It should be noted that since cosmic censorship is such a
hard problem it is common to consider the related problem where the initial
data is generic subject to some extra restrictions, e.g. symmetry 
restrictions. This is the case in the results of this paper.

The main theorem of the paper can now be given. It is a statement about 
strong cosmic censorship for the class of spacetimes involved.
  
\begin{theorem}\label{scc}
For data belonging to the open dense subset ${\cal G}_c$ of 
the set of smooth initial data for the Einstein-Maxwell equations with 
polarized Gowdy symmetry and constant area radius the corresponding maximal 
Cauchy development is inextendible.
\end{theorem}

\noindent
{\bf Proof} Let $(g,A)$ be initial data belonging to ${\cal G}_c$. By the 
results of Section \ref{singularity} the maximal Cauchy development of these 
data is inextendible towards the past since the Kretschmann scalar is 
unbounded along any inextendible past-directed causal geodesic. By the results 
of Section \ref{future} the maximal development is future geodesically 
complete and hence inextendible towards the future. This completes the proof.

In fact inextendibility towards the future already follows from global
existence to the future in areal time, as has been shown in 
\cite{dafermos05a}. It is nevertheless good that more has been proved about
the future asymptotics since this gives more insight into what is going on.
It should also be noted that the argument using global existence in areal
time is fundamentally restricted to spacetimes with symmetry while proving
geodesic completeness is a strategy which can in principle be applied to
situations without any symmetry. 

Note that the fact that the above theorem deals only with initial data with
constant area radius is not a serious restriction since any solution admits
a Cauchy hypersurface carrying data of this type.

In this paper the dynamics of solutions of the Einstein-Maxwell equations with
polarized Gowdy symmetry has been analysed rather comprehensively. This was 
based on the existing analysis of the dynamics of vacuum Gowdy spacetimes
which need not be polarized. This raises the obvious question, whether it 
is possible to do something similar for general solutions of the 
Einstein-Maxwell equations with Gowdy symmetry. Some aspects of this 
problem will now be discussed. In the full Einstein-Maxwell case there
are two metric coefficients $P$ and $Q$ as in the vacuum case and two 
components $\omega$ and $\chi$ of the potential.

In looking at the behaviour of the solutions in the past a possible starting 
point is to try to extend the results on the existence of low velocity 
solutions of the Gowdy equations proved in \cite{kr} using Fuchsian methods. 
A more complicated system of this type, the Einstein-Maxwell-dilaton-axion 
system was discussed from this point of view in \cite{narita00}. Setting the 
dilaton, the axion and the dilaton coupling constant in that system to zero 
gives rise to the Einstein-Maxwell system. A special case considered in
\cite{narita00} is in principle relevant to this but unfortunately it is
not self-consistent in the non-polarized case since one of the field 
equations was neglected (cf. the remarks on p. 166 of \cite{rendall08}).
Other special cases considered in \cite{narita00} may provide an avenue to
a greater understanding of singularities of solutions of the full 
Einstein-Maxwell-dilaton-axion system. In any case, the question of whether 
Fuchsian techniques can be used to obtain solutions of the Einstein-Maxwell 
equations with Gowdy symmetry depending on the full number of free functions 
(i.e. eight) remains open.

Global existence in the future in areal time is known and this suffices to 
prove the part of strong cosmic censorship which concerns the future 
evolution using the arguments of \cite{dafermos05a}. On the other hand 
the detailed asymptotics have not been determined. Parts of the argument 
which applies to the vacuum Gowdy case have been extended to the 
Einstein-Maxwell case in \cite{ringstrom06b}. These are the parts which 
concern the decay of an energy-like quantity analogous to those defined
in (\ref{energy1}) and (\ref{energy2}). However other important parts of 
what was done in the vacuum case have not yet been generalized.

These remarks show that the task of extending the results on the 
Einstein-Maxwell equations with polarized Gowdy symmetry obtained in this 
paper to the case of solutions with fully general Gowdy symmetry presents 
some interesting challenges. 

\vskip 1cm\noindent
{\bf Appendix} {\bf The dominant energy condition for the Maxwell field.}

It is common knowledge that the Maxwell field satisfies the dominant
energy condition but it is difficult to find a concise, elementary and 
self-contained proof in the literature. For the convenience of the reader
one will be included here. The dominant energy condition for an 
energy-momentum tensor $T^{\alpha\beta}$ can be formulated as the condition
that $T_{\alpha\beta}v^\alpha w^\beta\ge 0$ for all future-pointing causal
vectors $v^\alpha$ and $w^\beta$. Since any future-pointing causal vector
can be written as a sum of two future-pointing null vectors it suffices
to verify the condition in the case that $v^\alpha$ and $w^\alpha$ are equal
to future-pointing null vectors $l^\alpha$ and $n^\alpha$. By possibly
rescaling $l^\alpha$ and adding two spacelike vectors $x^\alpha$ and 
$y^\alpha$ these may be completed to a null basis where the vectors have the 
following inner products: $l_\alpha n^\alpha=-1$, 
$l^\alpha x_\alpha=l^\alpha y_\alpha=n^\alpha x_\alpha=n^\alpha y_\alpha=0$,
$x^\alpha x_\alpha=y^\alpha y_\alpha=1$, $x^\alpha y_\alpha=0$. The
electromagnetic field can be expanded in this basis to give
\begin{equation}
T_{\alpha\beta}=Al_{[\alpha}n_{\beta]}+Bl_{[\alpha}x_{\beta]}+
Cl_{[\alpha}y_{\beta]}+Dx_{[\alpha}y_{\beta]}
+En_{[\alpha}x_{\beta]}+Gn_{[\alpha}y_{\beta]}
\end{equation}  
where the coefficients are constants. Straightforward computations give
\begin{equation}
F_{\alpha\gamma}F_\beta{}^\gamma l^\alpha n^\beta=\frac14 (A^2+EB+GC),\ \ \ 
F_{\alpha\beta}F^{\alpha\beta}=\frac12 (D^2-A^2)-EB-GC
\end{equation}
and 
\begin{equation}
T_{\alpha\beta}l^\alpha n^\beta=\frac{1}{32\pi}(A^2+D^2).
\end{equation}


\begin{thebibliography}{12}

\bibitem{arr} Andr\'easson, H., Rein, G. and Rendall, A. D. 2003 On the
Einstein-Vlasov system with hyperbolic symmetry. Math. Proc. Camb. Phil.
Soc. 134, 529--549.
\bibitem{arw} Andr\'easson, H., Rendall, A. D. and Weaver, M. 2004 Existence 
of CMC and constant areal time foliations in $T^2$ symmetric spacetimes with
Vlasov matter. Commun. PDE 29, 237--262.
\bibitem{bcim} Berger, B., Chru\'sciel, P. T., Isenberg, J. and Moncrief, V.
1997 Global foliations of vacuum spacetimes with $T^2$ isometry. Ann.
Phys. (NY) 260, 117--148.
\bibitem{carmeli} Carmeli, M., Charach, Ch. and Malin, S. 1981 Survey of 
cosmological models with gravitational, scalar and electromagnetic waves.
Phys. Rep. 76, 79--156.
\bibitem{chrusciel90} Chru\'sciel, P. T., Isenberg, J. and Moncrief, V.
1990 Strong cosmic censorship in polarised Gowdy spacetimes. Class. Quantum 
Grav. 7, 1671--1680.
\bibitem{dafermos05a} Dafermos, M. and Rendall, A. D. 2006 Inextendibility of
expanding cosmological models with symmetry. Class. Quantum Grav. 22, 
L143--L147.
\bibitem{dafermos06a} Dafermos, M. and Rendall, A. D. 2006 Strong cosmic
censorship for $T^2$ symmetric cosmological spacetimes with collisionless
matter. Preprint gr-qc/0610075.
\bibitem{dafermos07a} Dafermos, M. and Rendall, A. D. 2007 Strong cosmic
censorship for surface-symmetric cosmological spacetimes with collisionless
matter. Preprint gr-qc/0701034.
\bibitem{kr} Kichenassamy, S. and Rendall, A. D. 1998 Analytic description 
of singularities in Gowdy spacetimes. Class. Quantum Grav. 15, 1339--1355.
\bibitem{moncrief81} Moncrief, V. 1981 Global properties of Gowdy spacetimes
with $T^3\times {\bf R}$ topology. Ann. Phys. (NY) 132, 87--107.
\bibitem{narita00} Narita, M., Torii, T. and Maeda, K. 2000 Asymptotic
singular behaviour of Gowdy spacetimes in string theory. Class. Quantum Grav
17, 4597--4613.
\bibitem{nungesser} Nungesser, E. 2008 
Strong cosmic censorship in polarized $T^3$-Gowdy symmetric spacetimes
with a Maxwell field. Diploma thesis, Free University, Berlin.
\bibitem{rendall95} Rendall, A. D. 1995 Global properties of locally spatially 
homogeneous cosmological models with matter. Math. Proc. Camb. Phil. Soc. 118, 
511--526.
\bibitem{rendall97a} Rendall, A. D. 1997 Existence of constant mean 
curvature foliations in spacetimes with two-dimensional local symmetry.
Commun. Math. Phys. 189, 145--164.
\bibitem{rendall08} Rendall, A. D. 2008 Partial differential equations in 
general relativity. Oxford University Press, Oxford.
\bibitem{ringstrom04} Ringstr\"om, H. 2004 On a wave map arising in
general relativity. Commun. Pure Appl. Math. 57, 657--703.
\bibitem{ringstrom06a} Ringstr\"om, H. 2006 Existence of an asymptotic 
velocity and implications for the asymptotic behaviour in the direction
of the singularity in $T^3$-Gowdy. Commun. Pure Appl. Math. 59,
977--1041.
\bibitem{ringstrom06b} Ringstr\"om, H. 2006 On the $T^3$-Gowdy symmetric 
Einstein--Maxwell equations. Ann. H. Poincar\'e 7, 1--20.
\bibitem{ringstromta} Ringstr\"om, H. Strong cosmic censorship in
$T^3$-Gowdy spacetimes. Preprint. To appear in Ann. Math. 
\bibitem{smulevici} Smulevici, J. 2008 Strong cosmic censorship for
$T^2$-symmetric spacetimes with positive cosmological constant and matter.
Ann. H. Poincar\'e 9, 1425--1453.
\bibitem{tegankong} Tegankong, D. 2005 Global existence and future 
asymptotic behaviour for solutions of the Einstein-Vlasov-scalar field
system with surface symmetry. Class. Quantum Grav. 22, 2381--2392.
\bibitem{weaverphd} Weaver, M. 1999 Asymptotic behaviour of solutions to
Einstein's equation. PhD thesis, University of Oregon.

\end{thebibliography}
\end{document}